\documentclass{ws-mpla}

\usepackage[super,compress]{cite}
\usepackage{graphicx}
\usepackage{epstopdf}
\usepackage{amsfonts}
\usepackage{amssymb}
\usepackage{epsfig}
\usepackage{hyperref}
\usepackage{color}

\def\de#1/de#2{\frac{\partial {#1}}{\partial {#2}}}

\newcommand{\ba}{\begin{eqnarray}}
\newcommand{\ea}{\end{eqnarray}}
\newcommand{\be}{\begin{equation}}
\newcommand{\ee}{\end{equation}}

\begin{document}


%
\catchline{}{}{}{}{}
%

\title{Evolution of matter perturbations in the context of cosmic slowing down}

\author{
D.~Koh Cuende,
\footnote{
\href{mailto:d.koh01@ufromail.cl}{d.koch01@ufromail.cl} 
}
G.~Panotopoulos,
\footnote{
\href{mailto:grigorios.panotopoulos@ufrontera.cl}{grigorios.panotopoulos@ufrontera.cl}
}
}

\address{
Departamento de Ciencias F{\'i}sicas, Universidad de La Frontera, \\ Casilla 54-D, 4811186 Temuco, Chile. 
}

\maketitle


\begin{abstract}
We investigate dynamical, time-evolving dark-energy models exhibiting phantom behaviour within Einstein gravity by analysing the evolution of matter perturbations. In particular, we consider five dark-energy parameterisations that predict a slowing down of the current cosmic acceleration. We examine the evolution of the growth index and its first derivative, as well as the combination parameter and the statefinder diagnostics, and compare their behaviour with that of the concordance $\Lambda$CDM model. For the parameter choices considered, the present-day values of the growth index and its first derivative are found to be compatible with ranges reported in previous phenomenological studies. The combination parameter is also found to differ from its $\Lambda$CDM value, while the statefinder parameters provide additional means of distinguishing the considered dynamical dark-energy scenarios from the concordance model. Our results illustrate the phenomenological signatures associated with these models and identify observables that may be useful for future observational tests. We stress, however, that the present analysis does not constitute a statistical model comparison or establish an observational preference for any of the considered dynamical dark-energy models over $\Lambda$CDM.
\end{abstract}


\ccode{}

\section{Introduction}\label{Intro}

The origin and nature of dark energy (DE), the component responsible for the current accelerated expansion of the Universe \cite{SN1,SN2,turner}, remains one of the most profound challenges in modern theoretical cosmology. It is well established that, within the framework of Einstein’s General Relativity \cite{GR}, a Universe composed solely of radiation and non-relativistic matter cannot undergo accelerated expansion. By contrast, a non-vanishing (and positive) cosmological constant \cite{einstein,carroll} provides the most economical model, yielding excellent overall agreement with a wide range of observational data.

\smallskip

The $\Lambda$CDM model, based on collisionless dark matter and a positive cosmological constant, is remarkably successful, yet it is not without shortcomings. At present, the community is confronted with two major challenges: the cosmological constant problem \cite{zeldovich,weinberg}, and a set of observational discrepancies commonly referred to as cosmological tensions. More specifically, a significant disagreement has emerged in the determination of the Hubble constant $H_0$. Measurements inferred from high-redshift cosmic microwave background (CMB) data differ from those obtained via local, low-redshift observations; see, for example, \cite{tension,tension1,tension2,tension3}. The value reported by the PLANCK Collaboration \cite{planck1,planck2}, $H_0 = (67-68)~\text{km/(Mpc sec)}$, is lower than that derived from local measurements, $H_0 = (73-74)~\text{km/(Mpc sec)}$ \cite{hubble,recent}. Furthermore, in the context of large-scale structure formation, the growth rate inferred from redshift-space distortion measurements appears to be lower than that predicted by Planck data \cite{Macaulay:2013swa,Basilakos:2017rgc}. This discrepancy is commonly known as the $\sigma_8$ tension \cite{DiValentino:2020vvd}, and may point towards the need for new physics \cite{newphysics,earlyDE,ben}. For instance, it has been suggested that a weaker effective Newton's constant within modified gravity frameworks could alleviate the $\sigma_8$ tension; see, for example, \cite{LP1,LP2,LP3,LP4,Panotopoulos:2021heb}. For recent reviews of the challenges faced by the $\Lambda$CDM model, see \cite{challenges1,challenges2}.

\smallskip

The results of a recent analysis reported earlier this year appear to point towards a time-evolving dark energy component that permits so-called phantom behaviour, in which the equation-of-state parameter enters the regime $w < -1$ \cite{DESI}; see also \cite{desi1,desi2,desi3,desi4,desi5}. Given the success of General Relativity, in the present work we investigate, within Einstein's gravity, five concrete dynamical and time-evolving dark-energy models characterised by an equation-of-state parameter that can enter the phantom regime, as suggested by the DESI results.

\smallskip

Owing to the well-known problems associated with the cosmological constant, it is perhaps unsurprising that a wide variety of dark energy models have been proposed and studied over the years as possible alternatives to the $\Lambda$CDM paradigm. Broadly speaking, these models can be classified into two main categories. On the one hand, there exists a class of models based on modified or alternative theories of gravity, in which additional correction terms to General Relativity become relevant at cosmological scales. On the other hand, a second class introduces a new dynamical field characterised by an equation-of-state (EoS) parameter $w < -1/3$. In the first category, often referred to as geometrical dark energy, one finds, for instance, $f(R)$ theories of gravity \cite{mod1,mod2,HS,starobinsky}, brane-world models \cite{langlois,maartens,dgp}, and scalar--tensor theories of gravity \cite{BD1,BD2,leandros,PR}. In the second category, known as dynamical dark energy, one encounters models such as quintessence \cite{DE1}, phantom \cite{DE2,DE3}, quintom \cite{DE4,DE5}, tachyonic models \cite{DE6}, and k-essence \cite{DE7}. For a comprehensive review of the dynamics of dark energy, see, for example, \cite{copeland}.

\smallskip

It is well known that many dark-energy models predict very similar expansion histories and, as a result, can remain consistent with current observational data. It therefore becomes important to investigate suitably defined diagnostic quantities that can reveal differences between competing models. In this context, one useful approach is to consider the so-called statefinder parameters, $r,s$, defined as follows \cite{Sahni:2002fz,Alam:2003sc}
\begin{eqnarray}
r & = & \frac{\dddot{a}}{a H^3} \\
s & = & \frac{r-1}{3 (q-\frac{1}{2})}
\end{eqnarray}
Here, an overdot denotes differentiation with respect to cosmic time $t$, $H = \dot{a}/a$ is the Hubble parameter, and $q = - \ddot{a}/(a H^2)$ is the deceleration parameter. It is straightforward to verify that, in the context of the $\Lambda$CDM model, the statefinder parameters take the constant values $r=1$ and $s=0$. These parameters can be computed within a given model and may evolve differently from one model to another. Moreover, their values could, in principle, be extracted from future observations \cite{SNAP1,SNAP2}. The statefinder diagnostic has been applied to a wide range of dark-energy models; for a partial list, see, for example, \cite{diagnostic1,diagnostic2,diagnostic3,diagnostic4,diagnostic5,diagnostic6,diagnostic7,diagnostic8,diagnostic9,diagnostic10} and references therein; see also \cite{cosmography1,cosmography2,cosmography3,cosmography4,cosmography5,cosmography6,cosmography7,cosmography8,cosmography9,cosmography10} for related work on cosmography.

\smallskip

Depending on the details of the underlying theory of gravity and/or the properties of the assumed dark-energy model, the evolution of linear matter perturbations may be affected in several ways. Even if two dark-energy models yield similar late-time accelerating expansion histories, they may still differ in the matter perturbations they produce \cite{staro,radouane}. This feature provides an additional and important means of investigating different dark-energy models at low redshifts (see, e.g., \cite{extra1,extra2,extra3,extra4}). It is therefore essential to characterise as accurately as possible the growth of matter perturbations. A quantity that has been extensively studied over the years is the so-called growth index, $\gamma$, introduced in \cite{gamma}. Indeed, the growth rate of matter perturbations can be probed by means of three-dimensional weak-lensing surveys \cite{Verde}.

\smallskip

In the present article, we extend the work of \cite{Magana:2014voa}, in which five different dark-energy parameterisations were studied and a slowing down of the current cosmic acceleration was reported. However, in \cite{Magana:2014voa} the statefinder parameters were not computed, and the evolution of matter perturbations was not investigated. Our aim here is to carry out this analysis, thereby providing a systematic comparison of the background and linear-growth properties of these five dynamical dark-energy parameterisations. Since the analysis involves phenomenological dark-energy models, the detailed evolution of the relevant cosmological quantities is necessarily model-dependent. Nevertheless, the comparison of the predicted behaviours over the redshift range considered provides useful information on the similarities and differences among the models. For constraints on the transition time in a model-independent way, see \cite{Muccino:2022rnd,Alfano:2024jqn}.

\smallskip

In addition, as we shall discuss later, the considered dynamical dark-energy models can lead to a different evolution of the combination parameter $A(z)$ compared with the concordance $\Lambda$CDM model. This comparison provides a phenomenological indication of how the models differ from the standard cosmological scenario and can be used to assess their consistency with the observational trends considered in this work. We stress, however, that the present analysis is not intended to establish a statistically preferred model. In particular, we do not perform a joint likelihood analysis, and do not consistently fit all relevant cosmological parameters, including the normalization of the matter power spectrum. Consequently, the comparison presented here should be understood as a study of the predicted cosmological evolution and its phenomenological consistency with the considered observational information, rather than as a statistical model-selection analysis.

\smallskip

The structure of this paper is as follows. After this introduction, in the next section we briefly review the standard Friedmann–Robertson–Walker Universe at the background level, as well as linear cosmological perturbation theory. In section 3 we present and discuss our main numerical results. Finally, we summarise our findings and offer concluding remarks in the last section. Throughout this work, we adopt the mostly positive metric signature, $(-,+,+,+)$, and use units in which the speed of light in vacuum is set to unity, $c = 1$.

\section{Theoretical framework}

The starting point is Einstein's GR \cite{GR} based on the Einstein-Hilbert term coupled to the matter content
\begin{equation}
S = \int d^4x \sqrt{-g} \left[ \frac{R}{16 \pi G}  + \mathcal{L}_M \right]
\end{equation}
where $g_{\mu \nu}$ is the metric tensor, $g$ is its determinant, $R$ is the corresponding Ricci scalar, $G$ is Newton's constant and $\mathcal{L}_M$ is the Lagrangian of the matter content. Varying the action with respect to the metric tensor
one obtains the well-known Einstein's field equations, which read
\begin{equation}
G_{\mu \nu} \equiv R_{\mu \nu} - \frac{1}{2} R g_{\mu \nu} = 8 \pi G \: T_{\mu \nu}
\end{equation}
where $R_{\mu \nu}$ is the Ricci tensor, while $T_{\mu \nu}$ is the matter energy-momentum tensor.

\subsection{Background evolution}

The basic cosmological equations governing the expansion of a homogeneous and isotropic Universe may be found e.g. in
\cite{review}. If matter consists of a perfect fluid with pressure $p$ and energy density $\rho$, the energy momentum 
tensor is given by
\begin{equation}
T_{\mu \nu} = p \: g_{\mu \nu} + (p + \rho) \: u_\mu \: u_\nu 
\end{equation}
where $u_\mu$ is the four-velocity of the fluid satisfying the condition $u_\mu u^\mu = -1$. The mixed component
stress-energy tensor takes the form \cite{review}
\begin{equation}
T_{\nu}^{\mu} =\textrm{diag}(-\rho,p,p,p)
\end{equation}

A spatially flat, isotropic and homogeneous universe is described by a Robertson-Walker metric \cite{review}
\begin{equation}
ds^2 = -dt^2 + a(t)^2 \delta_{ij} dx^i dx^j 
\end{equation}
where the scale factor $a(t)$ is the only unknown quantity, and all quantities depend on the cosmic time $t$ only.

\smallskip

The cosmological equations are found to be the continuity equation as well as the two Friedmann
equations \cite{review}
\begin{eqnarray}
H^2 & = & \frac{8 \pi G}{3} \rho \\
\frac{\ddot{a}}{a} & = & -\frac{4 \pi G}{3} (\rho + 3p) \\
0 & = & \dot{\rho} + 3 H (\rho+p)
\end{eqnarray}
where an over dot denotes differentiation with respect to cosmic time, and $H=\dot{a}/a$ is the Hubble parameter. The
second Friedmann equation may be written down equivalently as follows
\begin{equation}
\dot{H} = \frac{\ddot{a}}{a} - H^2 = - 4 \pi G (\rho+p)
\end{equation}
Finally, the deceleration parameter, $q$, is defined by
\begin{equation}
q \equiv - \frac{\ddot{a}}{aH^2}
\end{equation}
and as a function of red-shift, $z=-1+a_0/a$, with $a_0$ being the present value of the scale factor, it is computed to be
\begin{equation}
q(z) = -1 + (1+z) \frac{H'(z)}{H(z)},
\end{equation}
while the statefinder parameter $r$ versus red-shift is given by the following expression
\begin{equation}
r(z) =  (1+z) \: q'(z) + q(z) [1+2 q(z)].
\end{equation}
As far as the statefinders are concerned, prospects for resolving the Hubble constant tension with standard sirens were discussed for instance in \cite{Feeney:2018mkj}. The luminosity distance in Cosmology, $D_L(z)$, may be written down as a series in powers of red-shift $z$ as follows \cite{Feeney:2018mkj}
\begin{equation}
D_L(z) = \frac{z}{H_0} \left( 1 + \frac{1-q_0}{2} z - \frac{1-q_0-3 q_0^2 + j_0}{6} z^2+... \right)
\end{equation}
ignoring higher order terms, where the jerk parameter $j$ is one of the two statefinders, and the index 0 denotes present day values. We may determine the value of the Hubble constant $H_0$ if we can measure simultaneously both the red-shift and the luminosity distance, and then at higher $z$ the values $q_0,j_0$ as well.

In the following, instead of the cosmic time, $t$, we shall be using either the scale factor, $a$, or the red-shift, 
$z$, as the independent variable.

\smallskip

Next, if there are several non-interacting fluid components, then
\begin{eqnarray}
p & = & \sum_i p_i \\
\rho & = & \sum_i \rho_i \\
0 & = & \dot{\rho_i} + 3 H (\rho_i+p_i)
\end{eqnarray}
For barotropic fluids $p=w \rho$, where $w$ is the equation-of-state parameter. In the case of phantom DE, the equation-of-state parameter $w=p/\rho$ is lower than -1, which means that it violates the null energy condition, $p+\rho < 0$, leading to super acceleration, $\dot{H} > 0$, where both the Hubble parameter and the energy density increase with time. At a more fundamental level starting from a Lagrangian description, phantom DE may be modeled as a scalar field with the wrong sign in front of the kinetic term.

Finally, it is convenient to
introduce for each fluid component $X$ the normalized (dimensionless) density, $\Omega_X$, which is defined to be
\begin{align}
\Omega_X &\equiv \frac{\rho_{X}}{\rho_{c}}\,, \quad \rho_c=\frac{3H^2}{8 \pi G}\,,
\end{align}
and therefore the first Friedmann equation may be written down equivalently as a constraint of the following form
\begin{equation}
\sum_X \Omega_X = 1
\end{equation}
i.e. the normalized densities of all fluid components sum up to unity. 

In particular, the matter normalized density as a function of the scale factor takes the form
\begin{equation}
\Omega_m(a) = \frac{\Omega_{m,0}}{a^3 E(a)^2}
\end{equation}
where $\Omega_{m,0}$ is the matter normalized density evaluated at today, while the dimensionless Hubble parameter, $E(a)$, is given by 
\begin{align}
E(a)^2 \equiv \left(\frac{H(a)}{H_0}\right)^2 = \Omega_{m,0} a^{-3} + (1 - \Omega_{m,0}) F(a)
\end{align}
neglecting radiation at low red-shift, $1+z=a_0/a$, of order one or so, where $H_0$ is the Hubble parameter evaluated at today, and $F(a)$ is determined by the DE equation-of-state parameter as follows
\begin{equation}
w(z) = -1 + (1+z) \: \frac{F'(z)}{3 F(z)}.
\end{equation}
Clearly, the present value of both $E(a)$ and $F(a)$ is equal to unity
\begin{equation}
E(a=1) = 1, \; \; \; \; \; F(a=1) = 1.
\end{equation}

\subsection{Linear cosmological perturbations}

Let us briefly review linear cosmological perturbation theory within GR, see e.g. \cite{Mukhanov:2005sc,other}.

\smallskip

The goal is to solve the perturbed Einstein's field equations
\begin{equation}
\delta G_\nu^\mu = 8 \pi G \: \delta T_\nu^\mu
\end{equation}
When studying structure formation, only scalar perturbations are relevant. Regarding gravity, the metric tensor is given by
\begin{equation}
ds^2 = -(1 + 2 \Psi) dt^2 + (1-2 \Phi) \: a(t)^2 \: \delta_{ij} dx^i dx^j
\end{equation}
where $\Psi,\Phi$ are the Bardeen potentials. In the case of vanishing anisotropic stress, however, $\Phi=\Psi$, and therefore the line element takes the form 
\begin{equation}
ds^2 = -(1 + 2 \Psi) dt^2 + (1-2 \Psi) \: a(t)^2 \: \delta_{ij} dx^i dx^j
\end{equation}
with a single potential $\Psi$ both in the temporal and spatial sectors.

Regarding the cosmological fluid, the perturbed stress-energy tensor has the form
\begin{equation}
\delta T_0^0 = -\delta \rho, \qquad \delta T_j^i =  \delta p \: \delta_j^i
\end{equation}
The full set of coupled equations for matter and metric perturbations may be found e.g. in \cite{mariam,pano1,pano2}.

\smallskip

The Fourier transform of the density contrast, $\delta_k = \delta \rho_m / \rho_m$, with $k$ being the wave number,
for pressure-less matter satisfies the following linear differential equation \cite{rogerio,leandros2,review2}
\begin{equation}
\ddot{\delta_k} + 2H \dot{\delta_k} - 4 \pi G \rho_m \delta_k = 0
\end{equation}
at linear level, $\delta_k \ll 1$, and for sub-horizon scales, $k/(2 \pi) \gg a H$. Whether DE clusters or not can affect growth observables. In this article any possible effect of dark energy perturbations is neglected from the outset, i.e. we assume here that only non-relativistic matter clusters. Therefore, DE affects the evolution of matter perturbations through the Hubble parameter only.

\smallskip

During matter domination
\begin{equation}
a(t) \sim t^{2/3}, \; \; \; \; \; H(t) = \frac{2}{3 t}
\end{equation}
the matter density contrast  grows linearly with the scale factor, $\delta_k(a) \sim a$.

\smallskip

The equation for $\delta$ may take equivalently the following form 
\begin{align}
\delta''(a) + \left( \frac{3}{a} + \frac{E'(a)}{E(a)} \right) \delta'(a) - \frac{3}{2} \frac{\Omega_{m,0}}{a^5 E(a)^2} \delta(a) = 0
\end{align}
where for simplicity we drop the sub-index $k$, and a prime denotes differentiation with respect to the scale factor.

\smallskip

The growth index, $\gamma$, is defined to be
\begin{equation}
f = \Omega_m^\gamma, \; \; \; \; \; \;  f \equiv \frac{d \ln \delta}{d \ln a} = \frac{a}{\delta} \: \frac{d \delta}{d a}
\end{equation}
or equivalently
\begin{equation}
\gamma(a) = \frac{ln(f(a))}{ln(\Omega_m(a))}.
\end{equation}
For a constant DE equation-of-state, $w(z)=w$, there is an impressive agreement between the numerical result and an 
analytic approximation, and at lowest order one finds \cite{leandros2}
\begin{equation}
\gamma = \frac{3(w-1)}{6w-5}
\end{equation}
around $z \sim 1$. This reduces to $\gamma=6/11$ for $\Lambda$CDM ($w=-1$). In full generality, however, the growth index is a function of red-shift, with a non-vanishing derivative $\gamma'(z) \equiv d\gamma(z)/dz$. In an attempt to further improve on the analytic approximation, in \cite{radouane,leandros2} the authors considered an expansion to first order in $z$ of the form
\begin{equation}
\gamma(z) = \gamma_0 + \gamma_1 z
\end{equation}
which could have interesting observational consequences, and which is characterized by two parameters that may be identified with today's values of the functions $\gamma(z),\gamma'(z)$
\begin{equation}
\gamma_0 = \gamma(z=0), \;\;\;\;\;  \gamma_1 = \gamma'(z=0)
\end{equation}
The numerical values $\gamma_0, \gamma_1$ were constrained in \cite{leandros2}, and their allowed values were found to be
\begin{equation}
\gamma_0 = 0.77 \pm 0.29, \; \; \; \; \; \; \gamma_1 = -0.38 \pm 0.85
\end{equation}
or, fixing $\gamma_0=6/11$ and varying only $\gamma_1$ the allowed range was found to be
\begin{equation}
\gamma_1 = 0.17 \pm 0.54.
\end{equation}
For updated and more recent constraints on varying $\gamma(z)$ see \cite{Nesseris:2013, Basilakos:2013, Yin:2019, Mhamdi:2024}. It is worth mentioning that the results of \cite{Mhamdi:2024} are in good agreement with those of \cite{Nesseris:2013, Basilakos:2013}, which means that after 11 years the bounds were not dramatically shifted.

During matter domination, since $\delta \propto a$, by definition $f$ is found to be $f(a)=1$. Furthermore, the second order differential equation for $\delta$ may be written down equivalently as a first order differential equation for $f(a)$
as follows
\begin{equation}
a f'(a) + f(a)^2 + f(a) \left( 2 + a \frac{E'(a)}{E(a)} \right) = \frac{3 \Omega_m(a)}{2}.
\end{equation}

Finally, as far as the red-shift space distortion data are concerned, we need to compute the so-called combination parameter defined by
\begin{equation}
A(a) = f(a) \: \sigma(a), \; \; \; \; \; \; \sigma(a) = \sigma_8 \frac{\delta(a)}{\delta(a=1)}
\end{equation}
with $\sigma_8=\sigma(a=1)$ being the rms density fluctuation on scales $k_{\sigma_8}=0.125 \: h ~Mpc^{-1}$ \cite{mariam}. In this study we set $\sigma_8 = 0.8$ obtained recently in \cite{Franco:2024} in a robust, model-independent way without assuming a fiducial cosmology.

\section{Numerical results}

In this work we shall consider five DE parameterizations as follows:

{\bf Model 1} (Jassal-Bagla-Padmanabhan parameterization) \cite{JBP1, JBP2}: It is characterized by the following equation-of-state parameter
\begin{equation}
w(z) = w_0 + \frac{w_1 z}{(1+z)^2}.
\end{equation}
Therefore, the function $F(z)$ is computed to be
\begin{equation}
F(z) = (1+z)^{3 (1+w_0)} exp\left( \frac{3 w_1 z^2}{2 (1+z)^2} \right)
\end{equation}
{\bf Model 2} (Feng-Shen-Li-Li parameterization I) \cite{FSLL}:
It is characterized by the following equation-of-state parameter
\begin{equation}
w(z) = w_0 + \frac{w_1 z}{1+z^2}.
\end{equation}
Therefore, the function $F(z)$ is computed to be
\begin{equation}
F(z) = (1+z)^{3 (1+w_0)} exp\left( \frac{3 w_1 arctan(z)}{2} \right) (1+z^2)^{3 w_1/4} (1+z)^{-3 w_1/2}
\end{equation}
{\bf Model 3} (Feng-Shen-Li-Li parameterization II) \cite{FSLL}:
It is characterized by the following equation-of-state parameter
\begin{equation}
w(z) = w_0 + \frac{w_1 z^2}{1+z^2}.
\end{equation}
Therefore, the function $F(z)$ is computed to be
\begin{equation}
F(z) = (1+z)^{3 (1+w_0)} exp\left(- \frac{3 w_1 arctan(z)}{2} \right) (1+z^2)^{3 w_1/4} (1+z)^{3 w_1/2}
\end{equation}
{\bf Model 4} (Polynomial parameterization) \cite{SL}:
It is characterized by the following equation-of-state parameter
\begin{equation}
w(z) = -1+c_1 \frac{1+2 z}{1+z} + c_2 \left(  \frac{1+2 z}{1+z} \right)^2.
\end{equation}
Therefore, the function $F(z)$ is computed to be
\begin{equation}
F(z) = (1+z)^{3 (1-8 w_0+9 w_1)/2} exp\left( 3 z \frac{w_0 (52 z+40)-9 w_1 (5 z+4)+7 z+4}{8 (1+z)^2} \right) 
\end{equation}
where the coefficents $c_1,c_2$ are related to the values of $w(z)$ at $z=0$ and $z=0.5$ as follows \cite{Magana:2014voa}:
\begin{eqnarray} 
c_1 & = & \frac{16 w_0-9 w_{0.5}+7}{4} \\
c_2 & = & -3 w_0 + \frac{9 w_{0.5}-3}{4}
\end{eqnarray}
{\bf Model 5} (Barbosa-Alcaniz parameterization) \cite{BA}:
It is characterized by the following equation-of-state parameter
\begin{equation}
w(z) = w_0 + \frac{w_1 z (1+z)}{1+z^2}.
\end{equation}
Therefore, the function $F(z)$ is computed to be
\begin{equation}
F(z) = (1+z)^{3 (1+w_0)} (1+z^2)^{3 w_1/2}.
\end{equation}
All five models considered here are characterized by 3 free parameters, namely $w_0,w_1,\Omega_{m,0}$. Upon comparison to SN data sets, the extracted numerical values of the parameters are shown in Table 1 of \cite{Magana:2014voa}. In this study we shall consider the Constitution \cite{Const}, Union 2 \cite{U2} and LOSS \cite{LOSS} SN data sets.

\smallskip

In the figures below, we show, for each parameterisation, the equation-of-state parameter, the deceleration parameter, the two statefinder parameters and the growth index. Regarding the combination parameter $A(z)$, for each supernova compilation we display, in the same plot, the five curves corresponding to the different parameterisations considered in this work. For comparison, we also show the corresponding curve for the concordance $\Lambda$CDM model. Available observational measurements are included as well, using the data reported in Table II of \cite{Mhamdi:2024}.

\smallskip

Regarding the deceleration parameter $q(z)$, we reproduce the results of \cite{Magana:2014voa}, namely a transition from a decelerated to an accelerated phase of expansion, together with a slowing down of the current cosmic acceleration at very low redshifts. The detailed behaviour depends on the supernova compilation, since the numerical values of the model parameters differ between the datasets. At the background level, we additionally analyse the evolution of $w(z)$ and of the two statefinder parameters $r(z)$ and $s(z)$, which were not considered in \cite{Magana:2014voa}. The models investigated here are time-evolving dynamical dark-energy scenarios whose equation-of-state parameter can enter the phantom regime.

\smallskip

Concerning the evolution of matter perturbations, the growth index $\gamma(z)$ is found to vary slowly with redshift, with a slightly negative first derivative at very low redshifts. For all the parameter choices considered, the present-day values $\gamma(0)$ and $\gamma'(0)$ fall within the ranges reported in previous phenomenological studies \cite{Yin:2019,Mhamdi:2024}; the corresponding values are listed in Tables 1--5. In particular, the values of $\gamma(0)$ remain close to the $\Lambda$CDM value (6/11), whereas $\gamma'(0)$ is negative for all the cases considered.

\smallskip

The combination parameter $A(z)=f(z) \: \sigma(z)$ starts at approximately (0.40)--(0.45) at the present epoch, increases to a maximum and subsequently decreases monotonically with increasing redshift. This qualitative behaviour is also found in the concordance $\Lambda$CDM model. For the five dynamical dark-energy parameterisations considered here, however, the predicted curves lie below the corresponding $\Lambda$CDM curve over most of the redshift range shown, with the difference being particularly pronounced for the Constitution and Union 2 supernova compilations. This comparison illustrates the phenomenological differences between the considered dynamical dark-energy scenarios and the concordance model.

\smallskip

It is important to emphasise that the comparison presented here should not be interpreted as evidence that any of the dynamical dark-energy models provides a statistically better fit to the observations than $\Lambda$CDM. The present work does not perform a likelihood-based parameter estimation or a formal statistical model comparison, and the value of $\sigma_8$ is not independently fitted within each model. Consequently, the comparison with the observational measurements shown in the figures is qualitative and does not establish an observational preference for the dynamical dark-energy scenarios considered here.

\smallskip

A dedicated statistical analysis, including a consistent treatment of the relevant cosmological parameters and of $\sigma_8$, would be required to assess whether these models are favoured by current observations and whether they can contribute to resolving the cosmological $\sigma_8$ tension. Such an analysis lies beyond the scope of the present work. In addition, the perturbation analysis performed here assumes that dark energy does not cluster. If dark-energy perturbations are included, the matter and dark-energy perturbations must be treated as a coupled system, potentially modifying the growth history and the associated observables. These issues provide interesting directions for future investigation.


\begin{figure*}[ht]
\centering
\includegraphics[width=0.45\textwidth]{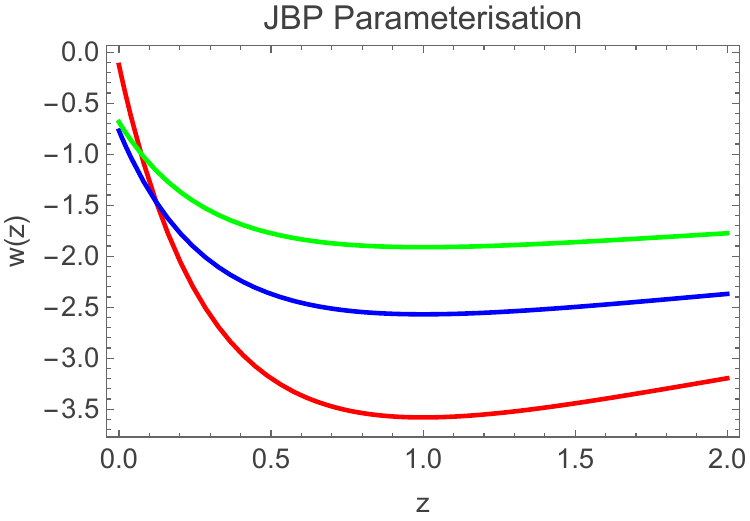}   \
\includegraphics[width=0.47\textwidth]{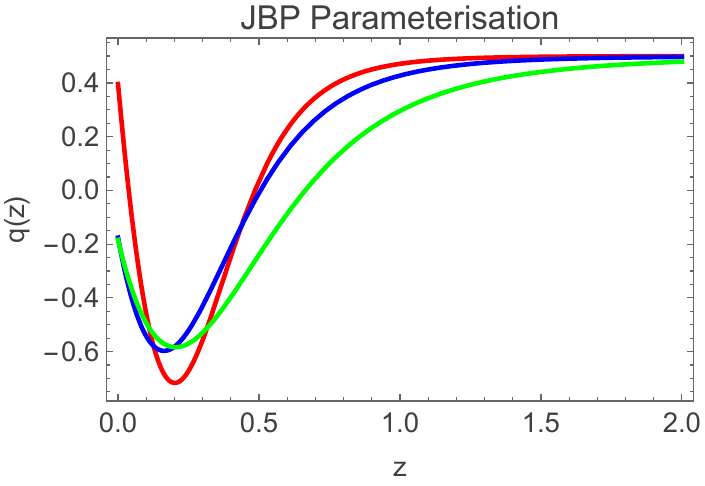}   
\caption{
{\bf Left:} Equation-of-state parameter, $w$, versus red-shift, $z$, for the JBP parameterization.
{\bf Right:} Deceleration parameter, $q$, versus red-shift, $z$, for the JBP parameterization. The different curves correspond to different SN compilation data as follows: Constitution (red), Union 2 (blue) and Loss-Union (green).
}
\label{fig:1}
\end{figure*}


\begin{figure}
\centering
\includegraphics[width=0.45\linewidth]{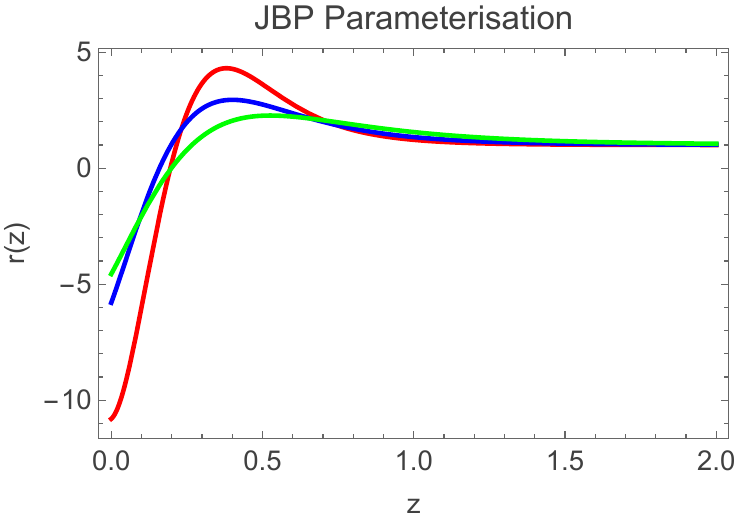} \
\includegraphics[width=0.45\linewidth]{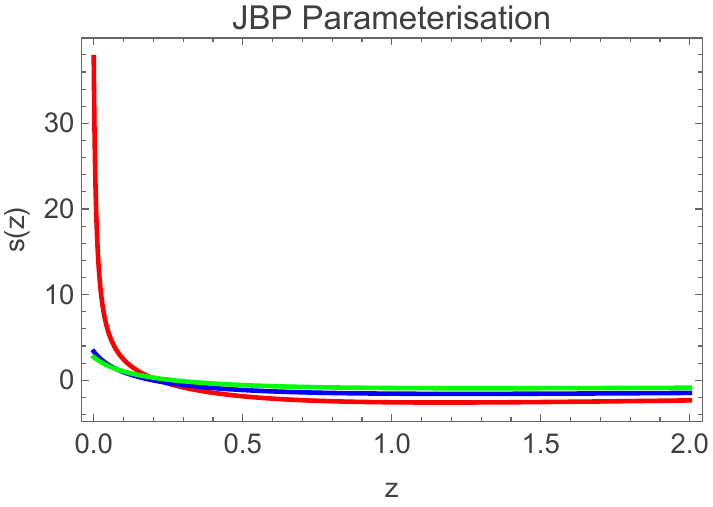} 
\caption{
{\bf Left:} First statefinder parameter, $r$, versus red-shift, $z$, for the JBP parameterization.
{\bf Right:} Second statefinder parameter, $s$, versus red-shift, $z$, for the JBP parameterization. The different curves correspond to different SN compilation data as follows: Constitution (red), Union 2 (blue) and Loss-Union (green).
}
\label{fig:2}
\end{figure}


\begin{figure}
\centering
\includegraphics[width=0.45\linewidth]{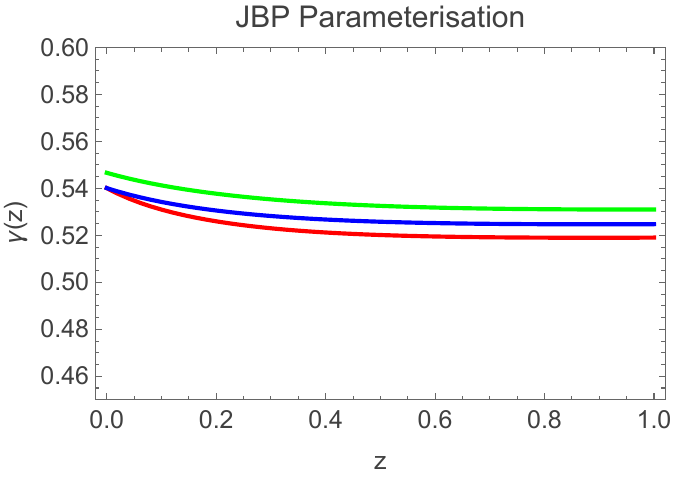} \
\includegraphics[width=0.47\linewidth]{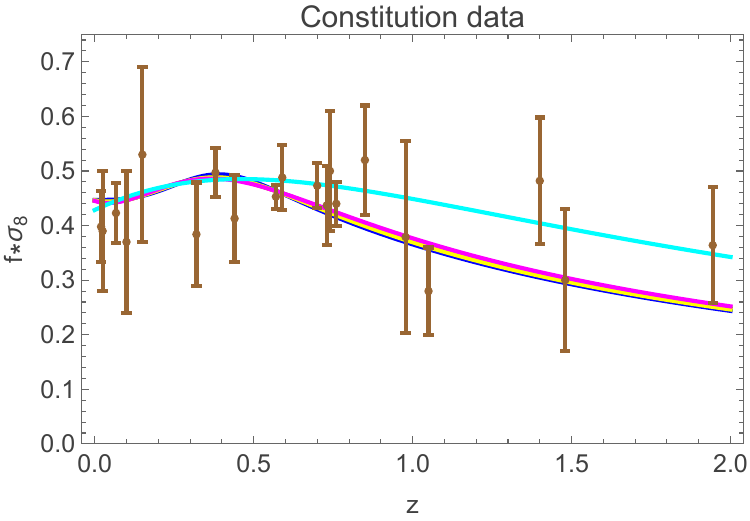} 
\caption{
{\bf Left:} Growth index, $\gamma$, versus red-shift, $z$, for the JBP parameterization.
{\bf Right:} Combination parameter, $A=f \sigma_8$, versus red-shift, $z$, for the Constitution data. The different curves correspond to different DE parameterisations. Observational points and the concordance model (in cyan) are shown as well.
}
\label{fig:3}
\end{figure}


\begin{figure*}[ht]
\centering
\includegraphics[width=0.45\textwidth]{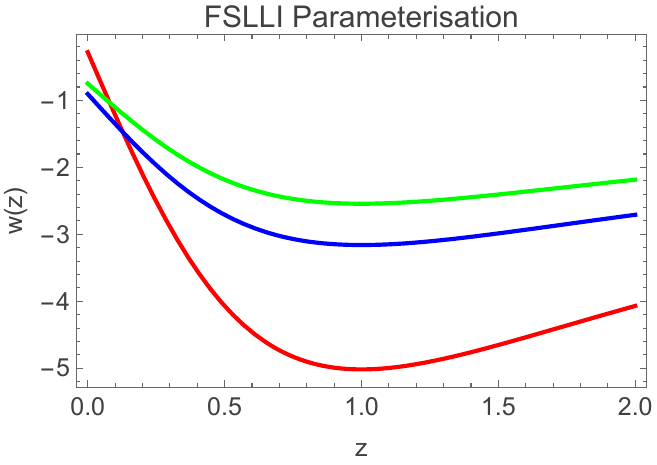}   \
\includegraphics[width=0.47\textwidth]{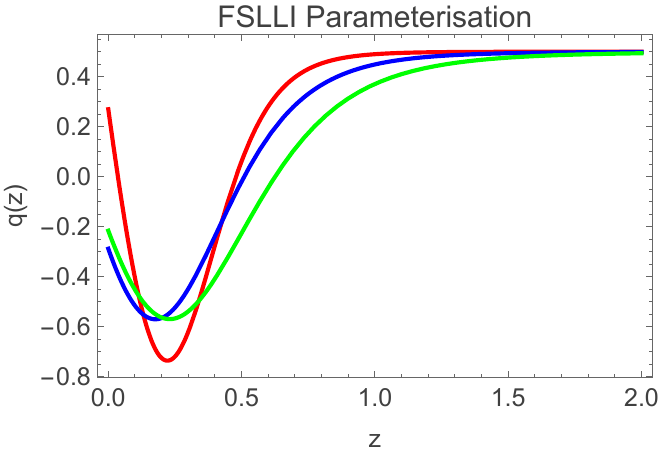}   
\caption{
{\bf Left:} Equation-of-state parameter, $w$, versus red-shift, $z$, for the FSLL I parameterization.
{\bf Right:} Deceleration parameter, $q$, versus red-shift, $z$, for the FSLL I parameterization. The different curves correspond to different SN compilation data as follows: Constitution (red), Union 2 (blue) and Loss-Union (green).
}
\label{fig:4}
\end{figure*}


\begin{figure}
\centering
\includegraphics[width=0.45\linewidth]{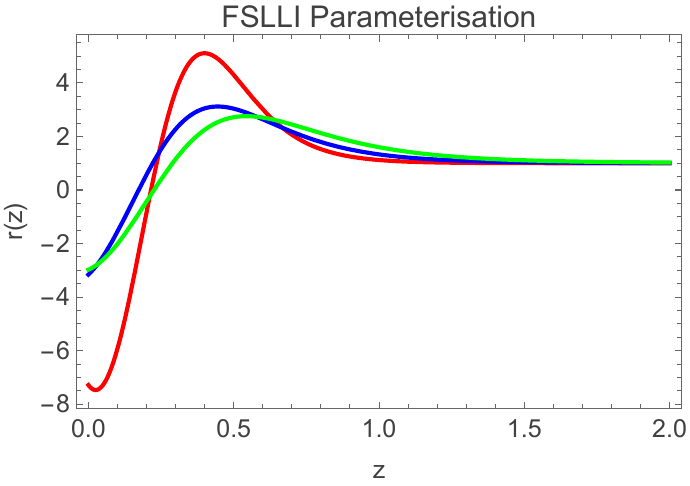} \
\includegraphics[width=0.45\linewidth]{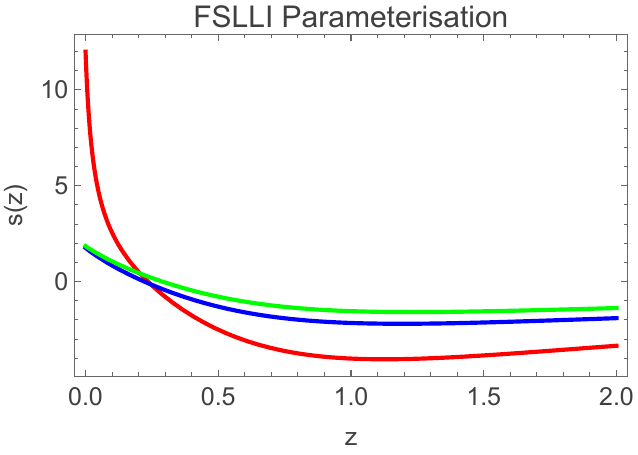} 
\caption{
{\bf Left:} First statefinder parameter, $r$, versus red-shift, $z$, for the FSLL I parameterization.
{\bf Right:} Second statefinder parameter, $s$, versus red-shift, $z$, for the FSLL I parameterization. The different curves correspond to different SN compilation data as follows: Constitution (red), Union 2 (blue) and Loss-Union (green).
}
\label{fig:5}
\end{figure}


\begin{figure}
\centering
\includegraphics[width=0.48\linewidth]{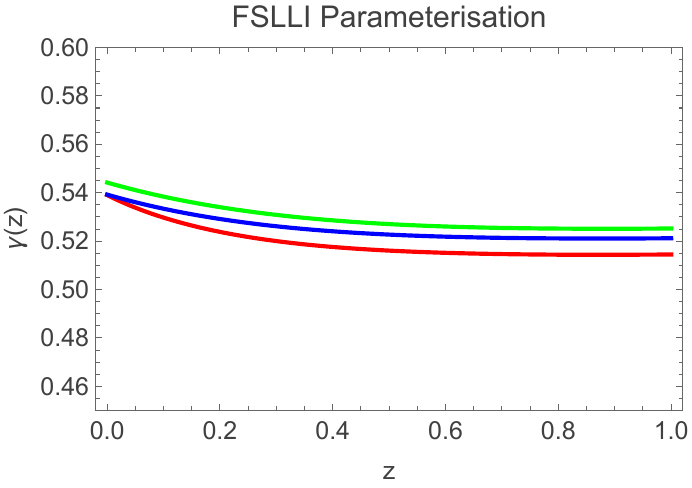} \
\includegraphics[width=0.47\linewidth]{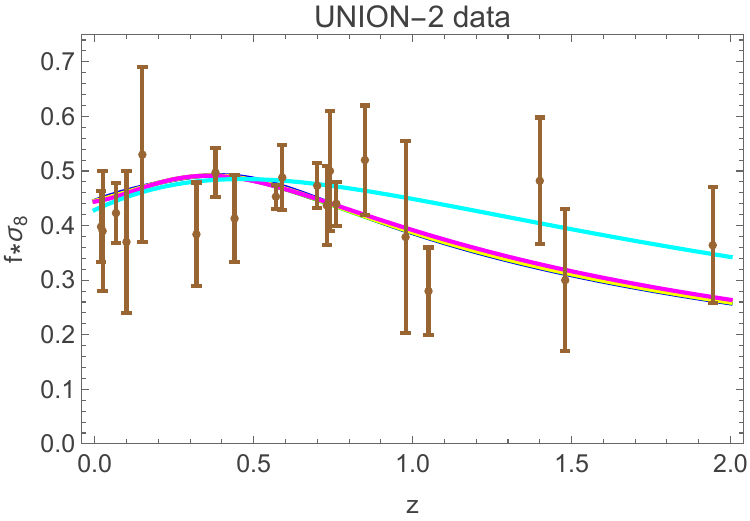} 
\caption{
{\bf Left:} Growth index, $\gamma$, versus red-shift, $z$, for the FSLL I parameterization.
{\bf Right:} Combination parameter, $A=f \sigma_8$, versus red-shift, $z$, for the Union 2 SN data. The different curves correspond to different DE parameterisations. Observational points and the concordance model (in cyan) are shown as well.
}
\label{fig:6}
\end{figure}


\begin{figure*}[ht]
\centering
\includegraphics[width=0.47\textwidth]{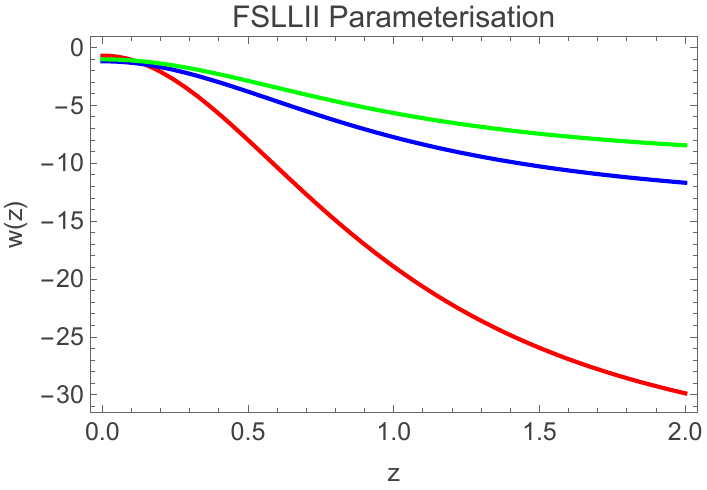}   \
\includegraphics[width=0.47\textwidth]{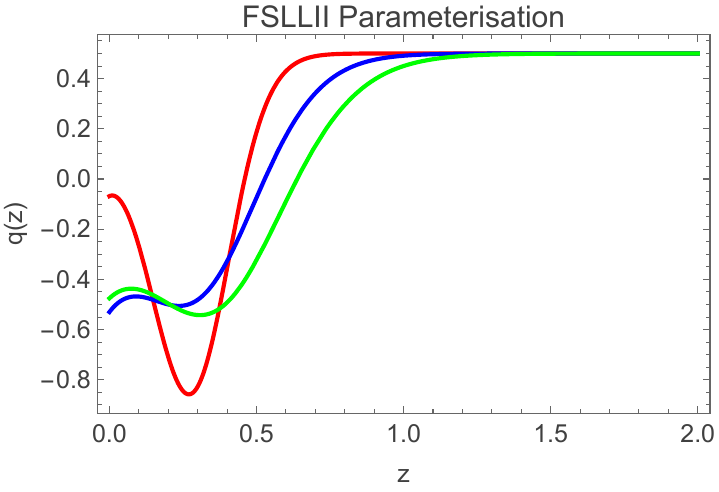}   
\caption{
{\bf Left:} Equation-of-state parameter, $w$, versus red-shift, $z$, for the FSLL II parameterization.
{\bf Right:} Deceleration parameter, $q$, versus red-shift, $z$, for the FSLL II parameterization. The different curves correspond to different SN compilation data as follows: Constitution (red), Union 2 (blue) and Loss-Union (green).
}
\label{fig:7}
\end{figure*}


\begin{figure}
\centering
\includegraphics[width=0.45\linewidth]{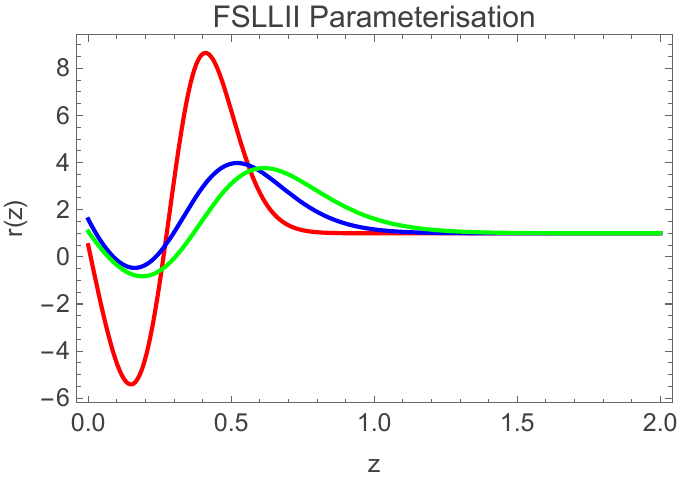} \
\includegraphics[width=0.45\linewidth]{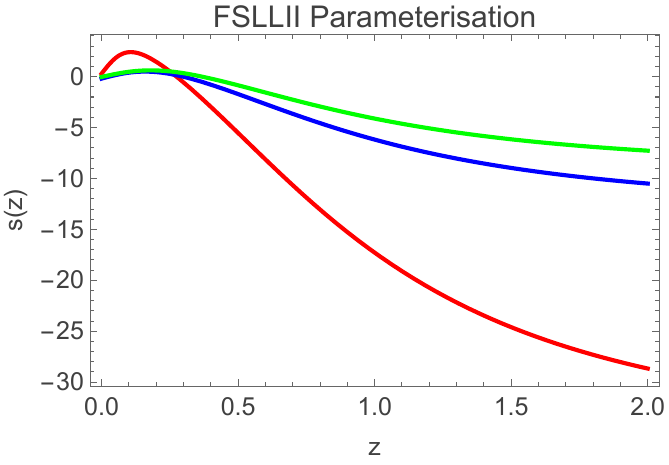} 
\caption{
{\bf Left:} First statefinder parameter, $r$, versus red-shift, $z$, for the FSLL II parameterization.
{\bf Right:} Second statefinder parameter, $s$, versus red-shift, $z$, for the FSLL II parameterization. The different curves correspond to different SN compilation data as follows: Constitution (red), Union 2 (blue) and Loss-Union (green).
}
\label{fig:8}
\end{figure}


\begin{figure}
\centering
\includegraphics[width=0.48\linewidth]{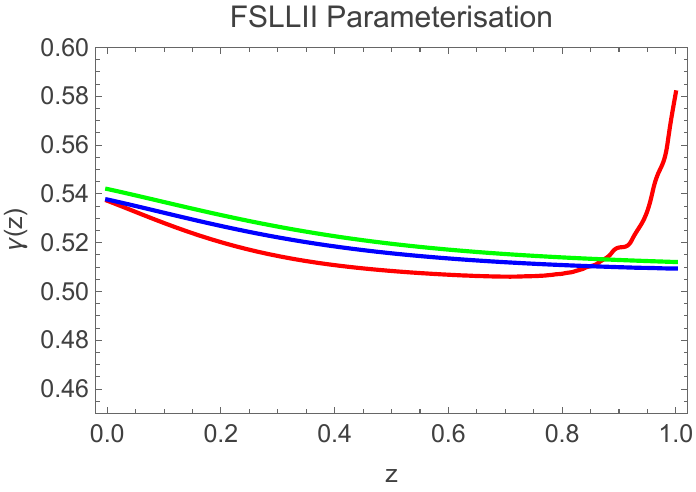} \
\includegraphics[width=0.47\linewidth]{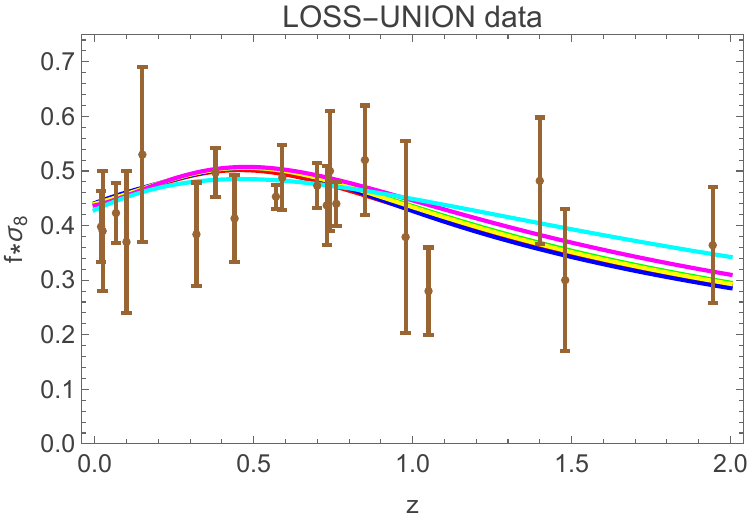} 
\caption{
{\bf Left:} Growth index, $\gamma$, versus red-shift, $z$, for the FSLL II parameterization.
{\bf Right:} Combination parameter, $A=f \sigma_8$, versus red-shift, $z$, for the LOSS-Union SN data. The different curves correspond to different DE parameterisations. Observational points and the concordance model (in cyan) are shown as well.
}
\label{fig:9}
\end{figure}


\begin{figure*}[ht]
\centering
\includegraphics[width=0.45\textwidth]{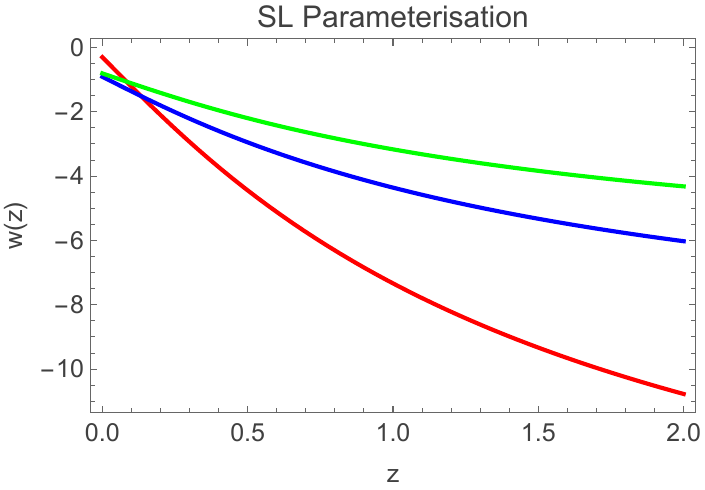}   \
\includegraphics[width=0.47\textwidth]{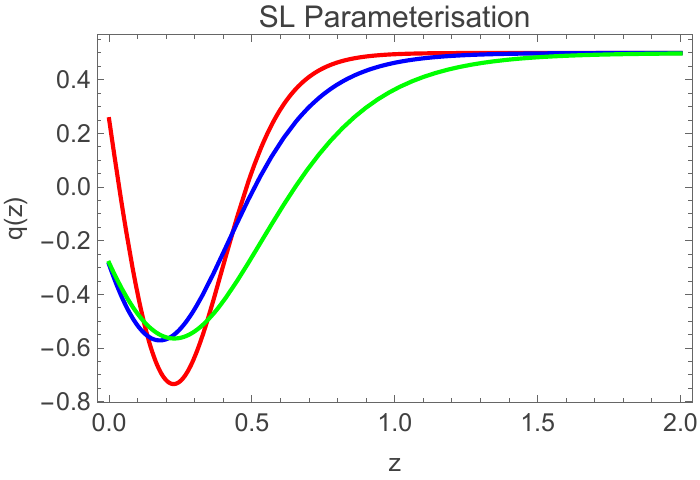}   
\caption{
{\bf Left:} Equation-of-state parameter, $w$, versus red-shift, $z$, for the polynomial parameterization.
{\bf Right:} Deceleration parameter, $q$, versus red-shift, $z$, for the polynomial parameterization. The different curves correspond to different SN compilation data as follows: Constitution (red), Union 2 (blue) and Loss-Union (green).
}
\label{fig:10}
\end{figure*}


\begin{figure}
\centering
\includegraphics[width=0.45\linewidth]{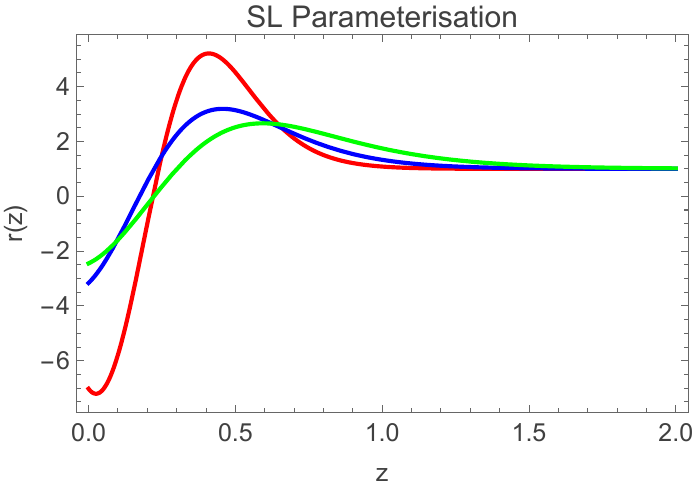} \
\includegraphics[width=0.45\linewidth]{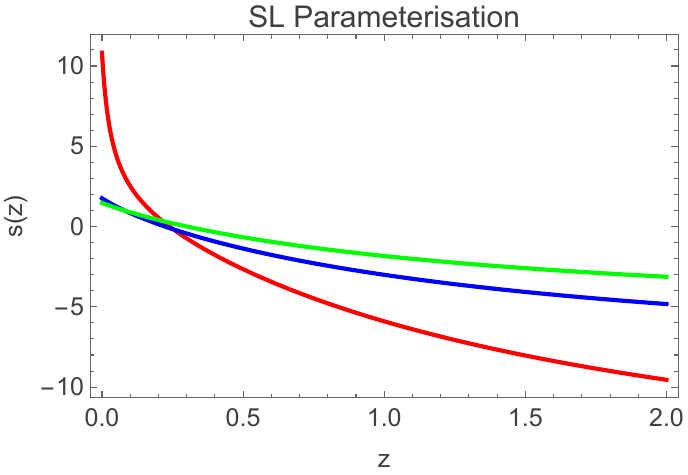} 
\caption{
{\bf Left:} First statefinder parameter, $r$, versus red-shift, $z$, for the polynomial parameterization.
{\bf Right:} Second statefinder parameter, $s$, versus red-shift, $z$, for the polynomial parameterization. The different curves correspond to different SN compilation data as follows: Constitution (red), Union 2 (blue) and Loss-Union (green).
}
\label{fig:11}
\end{figure}


\begin{figure}
\centering
\includegraphics[width=0.55\linewidth]{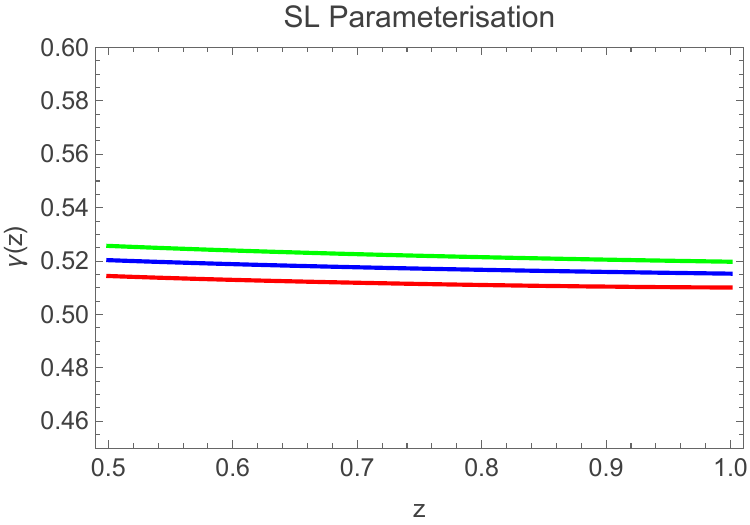} 
\caption{
Growth index, $\gamma$, versus red-shift, $z$, for the polynomial parameterization.
}
\label{fig:12}
\end{figure}


\begin{figure*}[ht]
\centering
\includegraphics[width=0.46\textwidth]{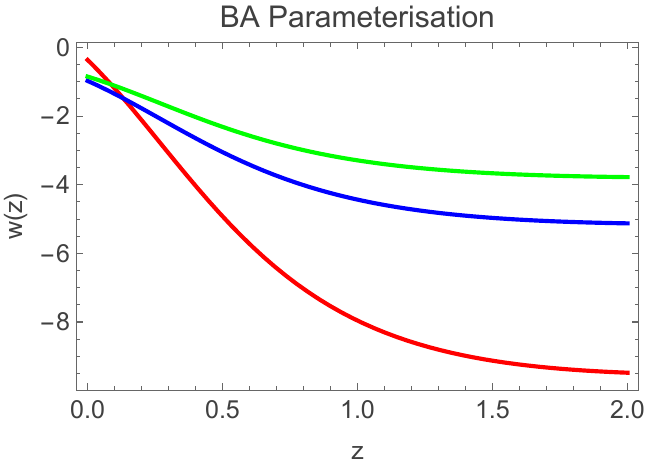}   \
\includegraphics[width=0.469\textwidth]{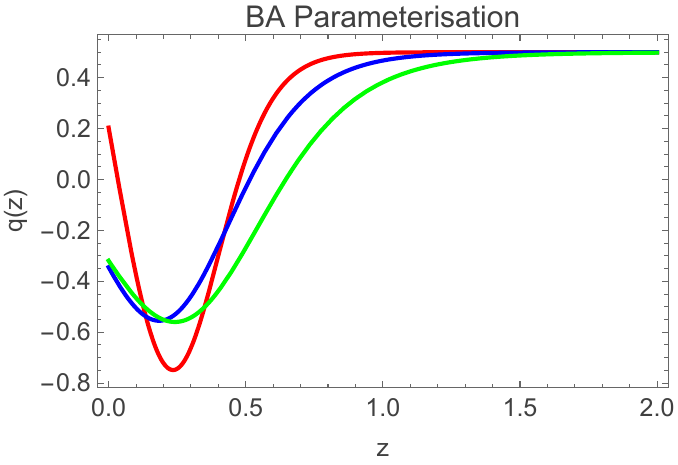}   
\caption{
{\bf Left:} Equation-of-state parameter, $w$, versus red-shift, $z$, for the BA parameterization.
{\bf Right:} Deceleration parameter, $q$, versus red-shift, $z$, for the BA parameterization. The different curves correspond to different SN compilation data as follows: Constitution (red), Union 2 (blue) and Loss-Union (green).
}
\label{fig:13}
\end{figure*}


\begin{figure}
\centering
\includegraphics[width=0.45\linewidth]{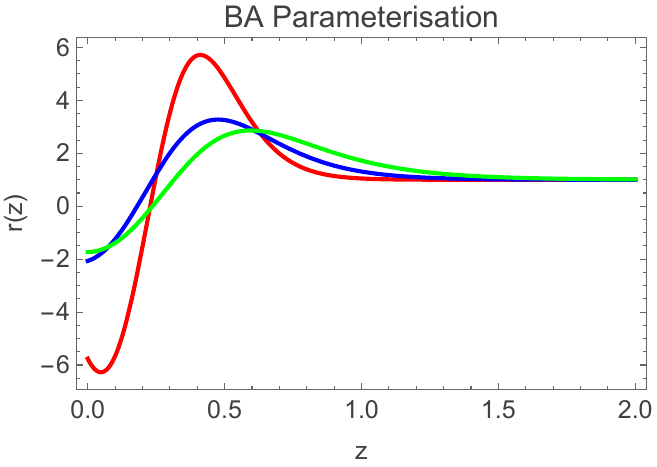} \
\includegraphics[width=0.45\linewidth]{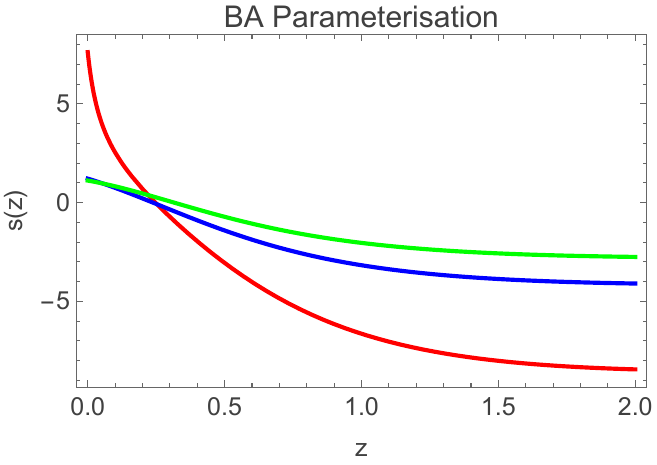} 
\caption{
{\bf Left:} First statefinder parameter, $r$, versus red-shift, $z$, for the BA parameterization.
{\bf Right:} Second statefinder parameter, $s$, versus red-shift, $z$, for the BA parameterization. The different curves correspond to different SN compilation data as follows: Constitution (red), Union 2 (blue) and Loss-Union (green).
}
\label{fig:14}
\end{figure}


\begin{figure}
\centering
\includegraphics[width=0.55\linewidth]{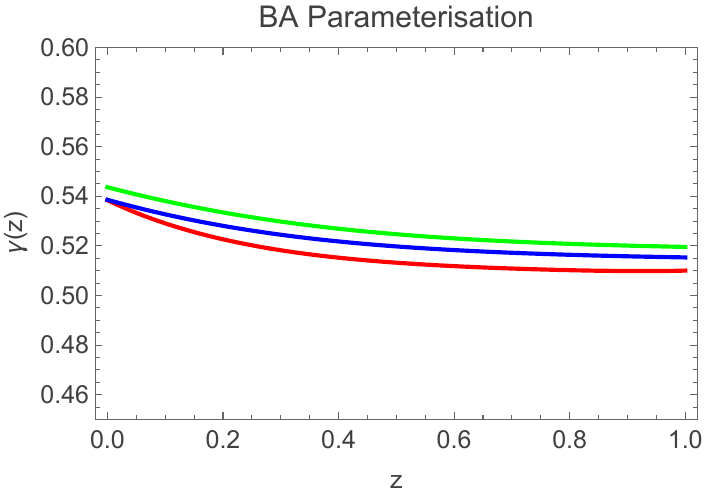} 
\caption{
Growth index, $\gamma$, versus red-shift, $z$, for the BA parameterization.
}
\label{fig:15}
\end{figure}



\begin{table}[]
\centering
\caption{
Values of $\gamma(0),\gamma'(0)$ for the JBP parameterisation
}
\label{tab:values1}
\begin{tabular}{@{}c|ccc@{}}
\hline
\hline
 & Constitution  & Union 2 & LOSS-Union 
\\
\hline
$\gamma(0)$ & 0.540 & 0.540 & 0.547 \\
\hline
$\gamma'(0)$  & -0.128 & -0.076 & -0.067
\end{tabular}
\end{table}


\begin{table}[]
\centering
\caption{
Values of $\gamma(0),\gamma'(0)$ for the FSLLI parameterisation
}
\label{tab:values1}
\begin{tabular}{@{}c|ccc@{}}
\hline
\hline
 & Constitution  & Union 2 & LOSS-Union 
\\
\hline
$\gamma(0)$ & 0.539 & 0.539 & 0.544 \\
\hline
$\gamma'(0)$  & -0.118 & -0.068 & -0.067 
\end{tabular}
\end{table}


\begin{table}[]
\centering
\caption{
Values of $\gamma(0),\gamma'(0)$ for the FSLLII parameterisation
}
\label{tab:values1}
\begin{tabular}{@{}c|ccc@{}}
\hline
\hline
 & Constitution  & Union 2 & LOSS-Union 
\\
\hline
$\gamma(0)$ & 0.537 & 0.538 & 0.542 \\
\hline
$\gamma'(0)$  & -0.089 & -0.051 & -0.051
\end{tabular}
\end{table}


\begin{table}[]
\centering
\caption{
Values of $\gamma(0),\gamma'(0)$ for the SL parameterisation
}
\label{tab:values1}
\begin{tabular}{@{}c|ccc@{}}
\hline
\hline
 & Constitution  & Union 2 & LOSS-Union 
\\
\hline
$\gamma(0)$ & 0.539 & 0.539 & 0.544 \\
\hline
$\gamma'(0)$  & -0.117 & -0.069 & -0.063
\end{tabular}
\end{table}


\begin{table}[]
\centering
\caption{
Values of $\gamma(0),\gamma'(0)$ for the BA parameterisation
}
\label{tab:values1}
\begin{tabular}{@{}c|ccc@{}}
\hline
\hline
 & Constitution  & Union 2 & LOSS-Union 
\\
\hline
$\gamma(0)$ & 0.538 & 0.539 & 0.544 \\
\hline
$\gamma'(0)$  & -0.113 & -0.065 & -0.061 
\end{tabular}
\end{table}


\section{Conclusions}

In summary, we have investigated dynamical dark-energy models within General Relativity using the statefinder diagnostic and the evolution of matter perturbations. We have considered five dark-energy parameterisations, namely JBP, FSLL I, FSLL II, SL and BA, and have examined their background evolution and their effects on the growth of matter perturbations.

At the background level, we have analysed the equation-of-state parameter, the deceleration parameter and the two statefinder parameters. At the level of linear cosmological perturbation theory, we have studied the growth index and the combination parameter. In the present analysis, dark-energy perturbations have been neglected by assuming that dark energy does not cluster. Consequently, dark energy affects the evolution of matter perturbations only through its contribution to the background expansion, as encoded in the Hubble parameter.

For the parameter choices considered, all five dynamical dark-energy models exhibit a transition from decelerated to accelerated expansion. Their evolving equation-of-state parameters can enter the phantom regime, and the models predict a slowing down of the current cosmic acceleration at very low redshifts. The growth index is found to vary slowly with redshift, and its present-day value and first derivative are compatible with ranges reported in previous phenomenological studies. The statefinder parameters further illustrate the differences between the dynamical dark-energy scenarios and the concordance $\Lambda$CDM model.

We have also found that the combination parameter is lower than its corresponding $\Lambda$CDM value for all five parameterisations and for the three supernova compilation datasets considered in this work. This result should be understood as a phenomenological difference between the predictions of the models and the concordance scenario, rather than as evidence for a statistically preferred model. In particular, the present analysis does not perform a likelihood-based parameter estimation or a formal statistical model comparison, and therefore does not establish an observational preference for any of the considered dynamical dark-energy models over $\Lambda$CDM.

Finally, the results presented here are subject to the assumption that dark energy remains homogeneous on the scales relevant to the matter perturbations. If dark energy perturbations are included, the matter and dark-energy perturbations must be treated as a coupled system, and the resulting growth history and associated observables may differ from those obtained here. A consistent treatment of dark-energy perturbations, together with a dedicated statistical analysis involving a likelihood-based fit, a consistent determination of $\sigma_8$ for each model, and appropriate model-comparison criteria, would therefore constitute an important direction for future work.

\section*{Acknowlegements}

We wish to thank the anonymous reviewer for a careful reading of the manuscript, for constructive criticism, as well as for useful comments and suggestions.





\end{document}